\documentclass[letterpaper,aastex]{emulateapj}
\usepackage{psfig}
\usepackage{graphicx}

\usepackage[usenames,dvips]{color}

\newcommand{\Fewbody}{{\em Fewbody}}

\begin{document}
\bibliographystyle{apj}

\title{On the Maximum Binary Fraction in Globular Cluster Cores}
                                                                                
\author{N.\ Ivanova, K. Belczynski\altaffilmark{1}, J.M.\ Fregeau\altaffilmark{2}, \& F.A.\ Rasio}
                                                                                
\affil{Northwestern University, Dept of Physics \& Astronomy,
       2145 Sheridan Rd, Evanston, IL 60208}
\altaffiltext{1}{Lindheimer Postdoctoral Fellow}
\altaffiltext{2}{MIT Department of Physics MIT}
                                                                                
\begin{abstract}{We study the evolution of binary stars in globular
clusters using a novel approach combining a state-of-the-art population synthesis
code with a simple treatment of dynamical interactions in the dense cluster core.
We find that the combination of stellar evolution and dynamical interactions
(binary--single and binary--binary) leads to a rapid depletion of the binary
population in the cluster core.
The {\em maximum\/} binary fraction today in the core of a typical dense cluster
like 47~Tuc, assuming an initial binary fraction of 100\%, is only about 5\%.
We show that this is in good agreement with recent {\em HST} observations of close
binaries in the core of 47~Tuc, provided that a realistic distribution of
binary periods is used to interpret the results. Our findings also have
important consequences for the dynamical modeling of globular clusters,
suggesting that ``realistic models'' should incorporate much larger
initial binary fractions than has usually been done in the past.
}
\end{abstract}
 

\section{Introduction}

Binary stars play a fundamental role in the dynamical evolution of
globular clusters, providing (through inelastic collisions) the source
of energy that supports them against gravothermal collapse
\citep{GoodmanHut_89,Gao_FP_91,2003ApJ...593..772F}. In the ``binary
burning'' phase, a cluster can remain in quasi-thermal equilibrium
with nearly constant core density and velocity dispersion for many
relaxation times, in much the same way that a star can maintain itself
in thermal equilibrium for many Kelvin-Helmholtz times by burning
hydrogen in its core.

At present, there are very few direct measurements of binary fractions
in clusters.  However, even early observations showed that binary
fractions in globular cluster cores are smaller than in the solar
neighborhood (e.g., \cite{Cote_M22_96}).  Recent Hubble Space
Telescope observations have provided further constraints
on the binary fractions in many globular clusters
\citep{2002AJ....123.2541B,1997ApJ...474..701R}.  The measured binary
fractions in dense cluster cores are found to be {\em very small\/}.
As an example, the upper limit on the core binary fraction of NGC 6397
is only 5-7\% \citep{CoolBolton_NGC6397_02}.

All dynamical interactions in dense cluster cores tend to {\em
destroy\/} binaries (with the possible exception of tidal captures,
which may form binaries, but turn out to play a negligible role; see
\S~3).  Soft binaries can be disrupted by any tidal interaction with
another passing star or binary.  Even hard binaries can be destroyed
in resonant binary--binary encounters, which typically eject two
single stars and leave only one binary remaining
\citep{1983MNRAS.203.1107M}, or produce physical stellar collisions
and mergers
\citep{1996MNRAS.281..830B,Fregeau_FB2_03}.

In addition, many binary stellar evolution processes can lead to
disruptions (e.g., following a supernova explosion of one of the
stars) or mergers (e.g., following a common envelope phase).  These
evolutionary destruction processes can also be enhanced by dynamics
(e.g., more common envelope systems form as a result of exchange
interactions; see Rasio, Pfahl, \& Rappaport 2000).

It is therefore natural to ask whether the small binary fractions
measured in old globular clusters today result from these many
destruction processes, and what the {\em initial binary fraction\/}
must have been to explain the current numbers. We address these
questions in this paper, by performing calculations that combine
binary star evolution with a treatment of dynamical interactions in
dense cluster cores.

\section{Methods and Assumptions}

Our initial conditions are described by the following parameters:
total number of stars (single or in a binary), $N$, initial mass function
(IMF), binary fraction, $f_b$, distribution of binary parameters
(period, $P$, eccentricity, $e$, and mass ratio, $q$). We typically
adopt standard choices used in population synthesis studies, which are
based on available observations for stars in the field and in young
star clusters (Sills et al. 2003).  For the calculations reported here, we
use the following initial conditions.  We adopt the IMF of
\cite{Kroupa_IMF_02}, which can be written as a broken power law
$dN\propto m^{-\alpha}dm$, where $\alpha=0.3$ for $0.01\le
m/M_\odot<0.08$, $\alpha=1.3$ for $0.08\le m/M_\odot<0.50$,
$\alpha=2.3$ for $m/M_\odot \ge 0.5$ We consider the mass range
$0.05\,M_\odot$ to $100\, M_\odot$.  The initial average stellar mass
is then $\langle m \rangle = 0.48\,M_\odot$.  The binary mass ratio,
$q$, is assumed to be distributed uniformly in the range $0 < q
<1$. This is in agreement with observations for $q \ga 0.2$
\citep{Woitas_MassRatio_01}.  The binary period, $P$, is taken from a
uniform distribution in $\log_{10} P$ over the range $P =
0.1$--$10^7\,$d.  The binary eccentricity, $e$, follows a thermal
distribution truncated such that there is no contact binary.

We evolve all stars (single and binary) using the population synthesis
code {\em StarTrack\/} \citep{Chris_02}. The evolution of single stars
is based on the analytic fits provided by \cite{Hurley_Single_00}, but
includes a more realistic determination of compact object masses
\citep{FryerKalogera_BH_01}. All our calculations use metallicity
$Z=0.001$, appropriate for a cluster such as 47~Tuc.  We treat the
evolution of stellar collision and binary merger products following
the prescription of \cite{Hurley_Binary_02}. To evolve the cluster
population, we consider two timesteps. One is associated with the
evolutionary changes in the stellar population, $\Delta t_{\rm ev}$,
and the other with the rate of encounters, $\Delta t_{\rm coll}$ (see
\S 3). $\Delta t_{\rm ev}$ is defined so that no more than 2\% of all
stars change their properties (mass and radius) by more than 5\%. The
global timestep for the cluster evolution is taken to be $\Delta t=
\min [t_{\rm ev},t_{\rm coll}]$

Our modeling of the cluster dynamics is highly simplified.  We assume
that the core number density, $n_{\rm c}$, and one-dimensional
velocity dispersion, $\sigma$, remain strictly constant throughout the
evolution.  These quantities are input parameters used to calculate
dynamical interaction rates in the cluster core (see below). While all
globular clusters have $\sigma \sim 10\,{\rm km}\,{\rm s}^{-1}$, the
core density can vary by several orders of magnitude. Here we set
$n_{\rm c}=10^5\,{\rm pc}^{-3}$ for most calculations, representative
of a fairly dense cluster like 47~Tuc. In general, $n_{\rm c}$ is the
main ``knob'' that we can turn to increase or decrease the importance
of dynamics.  Setting $n_{\rm c}=0$ corresponds to a traditional
population synthesis simulation, where all binaries and single stars
evolve in isolation after a single initial burst of star formation. To
model a specific cluster, we match its observed core luminosity volume
density $\rho^{\rm o}_L$, central velocity dispersion, and
half-mass relaxation time.

The escape speed from the cluster core can be estimated from
observations as $v_{\rm e} = 2.5\,\sigma_3$\citep{Webbink_GC_90}, where
$\sigma_3$ is the three-dimensional core velocity dispersion.
Following an interaction or a supernova explosion, any object that has
acquired a recoil speed exceeding $v_{\rm e}$ is removed from the
simulation.  For computing interactions in the core, the velocities of
all objects are assumed to be distributed according to a lowered
Maxwellian \citep{King_65}, with 
$f(v) = v^2/\sigma(m)^2 (\exp(-1.5 v^2/\sigma(m)^2) 
- \exp(-1.5 v_e^2/\sigma(m)^2))$ with parameters
$\sigma(m)= (\langle m \rangle/m)^{1/2}\sigma_3$ (assuming energy
equipartition in the core) and $v_{\rm e}$. In addition, we use
$\sigma$ to impose a cut-off for soft binaries entering the core.  Any
binary with maximum orbital speed $< 0.1 \sigma_3$ is immediately
broken into two single stars \citep{Hills_90}.

In the presence of a broad mass spectrum, the cluster core is always
dominated by the most massive objects in the cluster, which tend to
concentrate there via mass segregation.  As stars evolve, the {\em
composition\/} of the core will therefore change significantly over
time. To model mass segregation in our simulations, we assume that the
probability for an object of mass $m$ to enter the core after a time
$t_s$ follows a Poisson distribution,
$p(t_s)=(1/t_{sc})\exp(-t_s/t_{sc})$, where the characteristic
mass-segregation timescale is given by $t_{sc}=10 \left(\langle m
\rangle/m\right) t_{\rm rh}$ \citep{Fregeau_MS_02}. Here $t_{\rm rh}$
is the half-mass relaxation time, which we assume to be constant for a
given cluster.  We fix $\langle m \rangle=2 M_\odot$, as this value
gives, in our model, the best fit for the ratio of core mass to total
cluster mass in 47~Tuc.

All objects are allowed to have dynamical interactions after they have 
entered the cluster core. We use a simple Monte-Carlo prescription
to decide which pair of objects actually have an interaction
during each timestep.
We considerer separately binary--binary and binary--single interactions, 
as well as single--single encounters (tidal captures and collisions).
Tidal captures are treated using the approach described in \cite{Zwart_TC_93}.
If the pericenter distance is less than twice the sum of the stellar radii,
the encounter is treated as a physical collision and assumed to lead to a merger.
Each dynamical interaction involving a binary is calculated using
\Fewbody, a new numerical toolkit for simulating small-$N$ gravitational
dynamics that is particularly suited to performing 3-body and 4-body 
integrations \citep{Fregeau_FB2_03,Fregeau_FB1_03}.
A more detailed description of our
Monte-Carlo procedure will be given in \citep{Nata_2GCMethod_04}.

Initial conditions for all our reference models are given in
Table~1. All models have central velocity dispersion $\sigma_1=10{\rm
km}\,{\rm s}^{-1}$, and initial primordial binary fraction
$f_{b,0}=1$, except for Model~B05, which has $f_{b,0}=0.5$.  Our
assumed period distribution implies that about 60\% of primordial
binaries are hard.  The initial number of stars is $N = 2.5\times
10^5$, with the cluster core containing about 1\% of the stars
initially.

\section{Results}

\begin{deluxetable}{l l l  l l l  l l l}
\tabletypesize{\scriptsize}
\tablecaption{Reference models.}
\tablehead{
\colhead{Model}               & \colhead{$\log n_{\rm c}$}    &
\colhead{$\log t_{\rm rh}$}    & \colhead{$f_{\rm b,c}$}      & 
\colhead{$f_{0.5}$}          & \colhead{$f_{\rm wd}$} & \colhead{$f_{\rm b}$} }
\startdata
 1    & 5.0  & 9.0  & 0.08   &  0.07 & 0.04 & 0.65\\
 D3   & 3.0  & 9.0  & 0.21   &  0.25  & 0.10 &0.70\\
 D4   & 4.0  & 9.0  & 0.14   &  0.15  & 0.07 &0.66\\
 D6   & 6.0  & 9.0  & 0.03   &  0.02 & 0.02 &0.66\\
 T8   & 5.0  & 8.0  & 0.10   &  0.04  & 0.03 &0.22\\
 T10  & 5.0  & 10.  & 0.06   &  0.06  & 0.04 &0.85\\
 B05  & 5.0  & 9.0  & 0.05   &  0.05  & 0.05 &0.34\\
 \tableline
 ND   & 5.0 &  0.0  & 0.90   &  0.77 & 0.50 & 0.90
\enddata
\label{table}
\tablecomments{ $n_{\rm c}$ is the core number density in ${\rm pc}^{-3}$ (assumed fixed), 
$t_{\rm rh}$ is the half-mass relaxation time in yr,
$f_{\rm b,c}$ is the binary fraction in the core, $f_{0.5}$ is the binary fraction
for non-degenerate stars more massive than $0.5\,M_\odot$, 
$f_{\rm wd}$ is the binary fraction among white dwarfs and
$f_{\rm b}$ is the overall binary fraction in the cluster. 
ND is the model with no dynamical interactions (field population), where
all stars are assummed to be in the core from the beginning.
Values for all binary fractions are given at 14 Gyr.
}
\end{deluxetable}

First consider our results for a typical dense cluster, in Model~1.
Starting with 100\% binaries initially, the final core binary fraction
(at 14~Gyr), $f_{\rm b,c}$, is only 8\%.  This is strikingly low, given that the cluster started
with {\em all} binaries.
Decreasing the initial binary fraction, $f_{b,0}$, to a more reasonable (but still large) 50\%
reduces $f_{\rm b,c}$ further to 5\%, as shown in Model B05.
The dependence of $f_{\rm b,c}$ on $f_{b,0}$ is not linear. This is
mainly due to mass segregation: decreasing $f_{b,0}$ also increases the
ratio of mean binary mass to mean stellar mass in the cluster, thereby resulting 
in a higher concentration of binaries in the core.
The majority (about 75\%) of destroyed binaries were disrupted by close dynamical encounters
(or, rarely, following a supernova explosion). Note that some binaries that are initially hard
eventually become soft after undergoing significant mass loss due to stellar evolution.
About 20\% of the destroyed binaries experienced mergers, typically after significant hardening through interactions. 
The rate of binary destruction by mergers is about an order of magnitude higher 
in this model (Model~1) than for the corresponding field population (Model~ND). A few percent of 
the binaries lost were actually not destroyed but instead were ejected 
from the cluster as the result of strong encounters. 
Tidal capture did not play a significant role;
the total number of tidal capture binaries formed during the cluster lifetime is less than 1\% of the 
final number of binaries in the core. 
While the final core binary fraction is extremely low, the overall cluster 
binary fraction remains high, about 65\%, even after 14 Gyr (Fig.~1).
Note, however, that the surviving binaries include mainly
low-mass systems which never entered the core (about 70\% of the initial
binaries never entered the core and therefore never had a chance to interact).
The average primary mass among binaries remaining outside the core at 14~Gyr is 0.2 $M_\odot$.


\begin{figure} \plotone{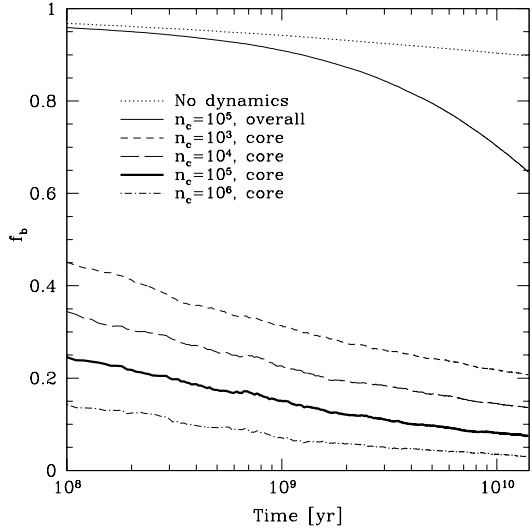} \caption{Evolution of the core and overall binary fraction in Model~1.  
Also shown is the evolution with no dynamics (ND) and for different core number densities 
(D3, D4 and D6).} \end{figure}  

Let us now compare results for different central densities, in Models 1, D3, D4 and D6. The evolution 
of $f_{\rm b, c}$ for these models is shown in Figure 1. For comparison, 
the binary fraction for the field case (Model ND) is  also presented. As expected, the core binary 
fraction decreases as $n_c$ increases.  
However, if one consider  the binary fraction for only non-degenerate stars,
its behavior is different. 
In particular, the final binary fraction for non-degenerate stars more massive 
than $0.5\,M_\odot$, $f_{0.5}$,  is higher than $f_{\rm b, c}$ for the low-density Model D3.  
Thus $f_{\rm b, c}$ is decreased partially through a lower binary fraction of degenerate objects. 
Degenerate objects, compared to non-degenerate  $0.5-0.9 M_\odot$ stars, 
evolved from initially more-massive stars and are more likely to have a more massive companion initially.
Their binary destruction rate is much higher, enchanced both by stellar evolution
(mass loss and mass transfer at more advanced evolutionary stages, 
SN explosions in a binary), and by dynamical interactions (large
cross-section for encounters). 


Next we examine how the half-mass relaxation time affects binary fractions  
(Models T8 and T10). We see that, surprisingly, the model with shorter relaxation time 
has a higher core binary fraction. 
There are two competing mechanisms that play a role here: mass segregation, 
which brings binaries into the core, and dynamical interactions, which destroy 
binaries in the core. 
A shorter segregation time increases the rate at which binaries enter the core but also 
allows less massive binaries  to interact. 
Therefore, the average mass of a binary in the  core will be smaller. 
However, the average time spent by a binary in the core also 
increases, so more can be destroyed, and the more massive binaries tend to be destroyed 
first as they have a larger interaction cross section. 
As a result, in Model~T8 $f_{\rm b,c}$ is 
higher than in T10, although the binary fraction of more massive binaries is  smaller. 


\begin{deluxetable}{l l l l l l l l}
\tabletypesize{\scriptsize}
\tablecaption{Models of specific clusters}
\tablehead{\colhead{Model}       & \colhead{$\log \rho^{\rm o}_{ L}$}
	& \colhead{$\log n_{\rm c}$}  & \colhead{$\sigma_1$}  
	& \colhead{$\log t_{\rm rh}$} & \colhead{$f_{b,c}$}   
	& \colhead{$f_{0.5}$}         & \colhead{$f_{\rm wd}$} }
\startdata
NGC3201& 2.73 \tablenotemark{a} \tablenotetext{a} {unless specified,  $\rho^{\rm o}_L$ 
and $t_{\rm rh}$ for specific clusters are taken from \cite{Harris_GCcatalog_96},
$\sigma_1$  from \cite{Pryor_GCdisp_93}.}
& 3.3 & 5.2 & 9.2    &  0.17  & 0.22 & 0.12 \\
$\omega$ Cen& 3.37 & 4.0 & 16  & 10    & 0.13 & 0.16 & 0.06 \\
M3  & 3.51 & 3.9  & 4.8 \tablenotemark{b} \tablenotetext{b}{\cite{Dubath_Disp_97}} & 8.5  & 0.16 & 0.13 & 0.07 \\
M4            & 3.82 &  4.3 & 4.2 & 8.82  & 0.12 & 0.10 & 0.07 \\
47Tuc  & 4.81 & 5.3 & 11.5 & 9.48 & 0.05 & 0.04 & 0.03 \\
NGC6397& 5.68 & 6.5 & 4.5 & 8.46  & 0.03 & 0.02 & 0.02 \\
\enddata
\label{table3}
\tablecomments{$\rho^{\rm o}_L$ is the observed core luminosity density in $L_\odot\,{\rm pc}^{-3}$,
$n_{\rm c}$ is the core number density in ${\rm pc}^{-3}$,
$\sigma_1$ is the velocity dispersion in ${\rm km}\,{\rm s}^{-1}$, 
$t_{\rm rh}$ is the half-mass relaxation time in yr,
$f_{\rm b,c}$ is the binary fraction in the core, $f_{0.5}$ is the binary fraction
for non-degenerate objects more massive than $0.5\ M_\odot$, and
$f_{\rm wd}$ is the ratio of WD in binaries to the total number of WD.
Values for all binary fractions are given at 14 Gyr.}
\end{deluxetable}
 
\begin{figure}
\plotone{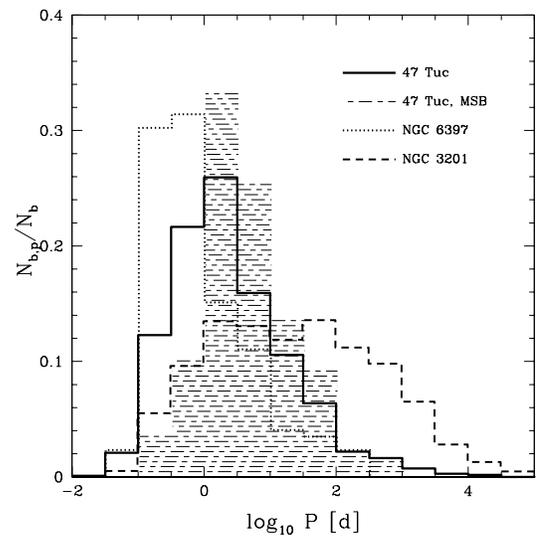}
\caption{The binary period distribution in the models of 
47 Tuc, $\omega$Cen, and NGC 6397 at 14 Gyr. 
The shaded histogram shows the period 
distribution of binaries containing two MS stars with
masses greater than $0.25\,M_\odot$ in 47~Tuc.
N$_{\rm b}$ is the total number of binaries and N$_{\rm b,p}$
is the number of binaries in the given period range.
}
\end{figure}

\section{Discussion and Comparison with Observations}

We performed several simulations with parameters that attempt to match those of specific globular 
clusters in the Galaxy (Table~2). In all cases the initial
binary fraction is 100\%, so our results for final core binary fractions represent 
upper limits.  As before we obtain very low values for $f_{\rm b,c}$.  
For example, in the 47~Tuc model, $f_{\rm b,c}$ is only 5\%.  At first glance, 
this may seem to conflict with 
observations. In particular, \cite{BinFreq_47Tuc_01} derive a binary fraction for the core 
of 47~Tuc of about 13\%, from observations of eclipsing binaries with periods
in the range $P\simeq 4$--$16\,$d. This estimate was based on an extrapolation
assuming a period distribution flat in $\log_{10} P$ from about $2\,$d to $50\,$yr. 
In Figure~2 we show the period distribution of core binaries in our simulation. 
Note that the period range of eclipsing binaries corresponds to the
peak of the distribution, while for larger $P$ it drops rapidly.
In particular, for binaries with components more massive than  0.25 $M_\odot$, 
the number of systems with periods in the range $2\,$d to $50\,$yr
is about 7 times smaller than would be predicted by a distribution flat in $\log_{10} P$. 
Using the observed number of eclipsing binaries and those from our simulation, the adjusted core
binary fraction from \citet{BinFreq_47Tuc_01} is about 4\%, which is consistent with our results. 
Figure~2 also shows the period distributions for models that represent the clusters NGC 3201 and NGC 6397. 
For denser clusters, such as NGC 6397, the peak of the distribution shifts toward
shorter periods, while for less dense clusters, such as NGC 3201, the distribution peaks at
longer periods and is flatter. We performed 3 additional simulations for 47~Tuc,
with $f_{\rm b,0}=0.25, 0.5 {\rm \ and \ } 0.75$. We found that with decreasing $f_{\rm b,0}$,
the adjusted core binary fraction decreases and reaches, e.g., 2.5\% for $f_{\rm b,0}=0.5$.

An alternative estimate of the binary fraction in the core of 47~Tuc is based on observations 
of BY~Dra stars \citep{BinFreq_47Tuc_01}.  Their estimated core binary fraction,
which can be considered a {\em lower limit}, is approximately 0.8\%, 
18 times lower than the estimate based on  eclipsing binaries. 
This estimate was based on 31 BY Dra binaries and 5 eclipsing binaries 
oberved in the period range 4-10 d. 
We analyzed the core binary population in our model in order to identify BY Dra binaries. 
We adopted the standard definition for a BY~Dra binary: primary mass in the range 
$0.3$--$0.7\,M_\odot$ (see, e.g., \cite{BoppFekel_BYDra_77})
and period in the range $4$--$10\,$d (as for the observed sample in 47~Tuc). 
The ratio between the total number of binaries and the number of BY~Dra systems is
found to be 38, 42 and 45 for models with $f_{\rm b,0}=1.0,0.75$ and 0.5, respectivly.
With 36 binaries observed in this period range, the total core binary fraction
is 2.9\%, 3.3\% and 3.5\% for $f_{\rm b,0}=1.0,0.75$ and 0.5, respectively.  
Based on these results, we estimate that the initial binary fraction should be 
at least 0.5, and more likely in the range $f_{\rm b,0}\simeq 0.75-1$.

Even more extreme results are obtained for NGC 6397.  This cluster
is classified observationally as ``core collapsed'' and therefore
may not be well described by our simplified dynamical model. Nevertheless, it is useful 
to study how the binary fraction in the  cluster would have evolved if the very high central 
density had been constant 
throughout the evolution. We find that the binary fraction 
for stars more massive than $0.5\,M_\odot$ is extremely low: 2\% in the core 
at 14~Gyr with 100\% binaries initially. For this cluster, there is a firm upper limit of 
3\% on the core binary fraction for stars in the mass range $0.45-0.8\,M_\odot$ and for binary mass ratios
$q>0.45$ \citep{CoolBolton_NGC6397_02}. For the same ranges in our model we find 2\%, in 
surprisingly good agreement with the observations.

In addition to their implications for the interpretation of observed binary fractions
in cluster cores, our results also have important consequences for the
theoretical modeling of globular clusters. Indeed, it is clear that ``realistic'' dynamical 
simulations of globular cluster evolution should include large populations of
primordial binaries, with initial binary fractions in the range $\sim 50\%$--$100\%$
(similar to what is usually assumed for the field; see, e.g., \cite{Multiple_91}). This poses a 
particular challenge for direct $N$-body simulations, where the treatment of even
relatively small numbers of binaries can add enormous computational costs.  For this
reason, current direct $N$-body simulations of star clusters with large initial
binary fractions typically have $N$ too small to be considered representative of
globular clusters (see, e.g., Portegies Zwart et al. 2003; Wilkinson et al. 2003).
However, approximate methods, such as Monte-Carlo, do not suffer from the same
limitations, and routinely simulate clusters with reasonably large $N$ 
($\sim 3\times 10^5$) and binary fraction ($\sim 30\%$), yet have not yet 
included advanced treatments of binary star evolution 
\citep[see, e.g., ][and references therein]{2003ApJ...593..772F}.

\nocite{Modest2}
\nocite{ecology_v}
\nocite{Wilkinson_2003}
\bibliography{liter} 

\end{document}